\documentclass[twocolumn,aps,showpacs,prb]{revtex4-1}
\usepackage{graphicx}
\usepackage{epstopdf}
\usepackage{amsmath}
\usepackage{multirow}

\begin{document}

\title{First-principles study of magnetic frustration in FeSe epitaxial films on SrTiO$_3$}
\author{Kai Liu$^{1,2}$}\email{kliu@ruc.edu.cn}
\author{Bing-Jing Zhang$^{1,2}$}
\author{Zhong-Yi Lu$^{1,2}$}\email{zlu@ruc.edu.cn}

\affiliation{$^{1}$Department of Physics, Renmin University of China, Beijing 100872, China}
\affiliation{$^{2}$Beijing Key Laboratory of Opto-electronic Functional Materials $\&$ Micro-nano Devices, Renmin University of China, Beijing 100872, China}

\date{\today}

\begin{abstract}

The effects of electron doping and phonon vibrations on the magnetic properties of monolayer and bilayer FeSe epitaxial films on SrTiO$_3$ have been studied, respectively, using first-principles calculations with van der Waals correction. For monolayer FeSe epitaxial film, the combined effect of electron doping and phonon vibrations readily leads to magnetic frustration between the collinear antiferromagnetic state and the checkerboard antiferromagnetic N\'eel state. For bilayer FeSe epitaxial film, such magnetic frustration is much more easily induced by electron doping in its bottom layer than its top layer. The underlying physics is that the doped electrons are accumulated at the interface between the FeSe layers and the substrate. These results are consistent with existing experimental studies.

\end{abstract}


\maketitle

\section{INTRODUCTION}

Iron-based superconductors have drawn worldwide interests both experimentally and theoretically since their discoveries.\cite{1111,122,111,11} The superconducting transition temperature $T_c$ can be modulated by charge doping via chemical substitution or by high pressure. In addition, epitaxial film growth is another effective way to tune the lattice parameters, magnetic properties, and $T_c$ of Fe-based superconductors.\cite{caolx10,xuePRB11,xueCPL12} By using molecular beam epitaxy (MBE) technique, atomically flat FeSe epitaxial films have been grown on bilayer graphene\cite{xuePRB11} and SrTiO$_3$ substrates.\cite{xueCPL12} Surprisingly, the monolayer FeSe on SrTiO$_3$ shows a superconducting transition signature around 77 K, exceeding the highest $T_c$ of bulk ferropnictides,\cite{renza08} while the bilayer FeSe does not exhibit a superconducting gap in the scanning tunneling microscope (STM) spectra.\cite{xueCPL12} With larger tensile strains on FeSe epitaxial films, the $T_c$ can also be lifted to $\sim$ 70 K.\cite{fengdlPRL14,fengdl14} Due to its simplest atomic structure among Fe-based superconductors, these fresh experiments have demonstrated that FeSe epitaxial film is a good model system for exploring the unconventional superconducting mechanism in Fe-based superconductors.

Various experiments have been carried out to investigate the physical properties of FeSe epitaxial films on SrTiO$_3$. \cite{zhouxjNC12,zhouxjNM13,fengdlNM13,xueAPE13,xueCPL14,maxc14,chucw13,shenzx13,zhouxj14a,zhouxj14b,jiajf14} Concerned with the electronic structure of FeSe monolayer on SrTiO$_3$, only electron-like Fermi surfaces exist at the corners of the Brillouin zone (BZ) in the angle resolved photoemission spectra (ARPES) measurement.\cite{zhouxjNC12,zhouxjNM13,fengdlNM13} This is quite different from that of multilayer FeSe films and bulk FeSe. It has been suggested that charge transfer from the substrate to FeSe layers plays an important role in the superconductivity. \cite{xueAPE13,xueCPL14,maxc14} For both monolayer and bilayer FeSe, more and more electrons are transferred from the substrate when annealing time increases.\cite{zhouxj14b} As for the magnetism, it was demonstrated that superconductivity occurs when the electron transfer from the substrate suppresses the otherwise pronounced spin density waves in monolayer FeSe.\cite{fengdlNM13} Recently, replica bands with a distance of 100 meV in ARPES measurement were found for monolayer FeSe epitaxial film.\cite{shenzx13} It was then proposed that this is induced by the coupling between electrons and substrate phonons, which enhances the superconducting pairing temperature profoundly.\cite{shenzx13} The above intensive experiments have provided valuable information about the FeSe epitaxial system.

Many efforts have also been devoted from a theoretical standpoint.\cite{liu12,leedh12,cohen13,zhangp13,zhangsb13,kuw14,gongxg14,duanwh14,xingdy14} In our previous calculations,\cite{liu12} the similar electronic structures of monolayer and bilayer FeSe on undoped SrTiO$_3$ suggested that the superconductivity would occur at the first layer of FeSe epitaxial film or at the interface. Xiang \textit{et al.} studied the influence of the screening effect of a ferroelectric phonon of the substrate on the Cooper pairs in the FeSe layer.\cite{leedh12} The electron doping effect has also been investigated. \cite{cohen13,zhangp13,zhangsb13} The measured Fermi surface of FeSe monolayer can be reproduced in calculation when the antiferromagnetic (AFM) N\'eel order is set up. \cite{cohen13,zhangp13} Cao \textit{et al.} have studied the effects of tensile strain and charge transfer on the spin density wave in FeSe/SrTiO$_3$ thin films.\cite{gongxg14} Li \textit{et al.} have calculated the electron-phonon coupling constant of FeSe/SrTiO$_3$ using the first-principles approach, indicating that the electron-phonon mechanism alone cannot explain its high $T_c$.\cite{xingdy14} The above theoretical works have proposed a variety of possible contributions to the superconductivity in FeSe epitaxial film from different points of view. Nevertheless, the physical mechanism of superconductivity in this system is still an open question. In particular, is the high transition temperature correlated with magnetism? Is there any other substantial phonon effect if the electron-phonon coupling alone cannot account for the high $T_c$? The entanglements of tensile lattice strain, electron doping, phonon, and magnetism in the FeSe epitaxial film further complicate the situation.

In this paper, we have studied the effects of electron doping and phonon vibrations on the magnetic properties of FeSe monolayer and bilayer on SrTiO$_3$ by using first-principles calculations with van der Waals (vdW) correction. The nonmagnetic, checkerboard AFM N\'eel, and collinear AFM states of FeSe epitaxial films have been investigated. The relationships of AFM fluctuations with electron doping and quantum zero-point (ZP) atomic displacements of phonon modes have been examined. The differences between monolayer and bilayer FeSe epitaxial films have also been addressed.

The rest of this paper is organized as follows. In Sec. II, the computational details are described. In Sec. III, the AFM variations of FeSe epitaxial films on SrTiO$_3$ have been studied as functions of electron doping and ZP atomic displacements. Discussions with related experimental and theoretical works are given in Sec. IV, and a short summary is provided in Sec. V.

\section{COMPUTATIONAL DETAILS}

Fully spin-polarized first-principles calculations were carried out by using the projector augmented wave (PAW) method\cite{paw} as implemented in the Vienna ab-initio simulation package.\cite{vasp} The generalized gradient approximation (GGA) of Perdew-Burke-Ernzerhof\cite{pbe} for the exchange-correlation potentials was adopted. To describe the vdW interaction in layered systems not included in the conventional density functional theory, the vdW-optB86b functional\cite{optb86} was chosen. To model FeSe ultra-thin films on TiO$_2$-terminated SrTiO$_3$(001) as grown in experiment,\cite{xueCPL12} we used a six-layer SrTiO$_3$(001) slab with FeSe monolayer and bilayer adsorbed on the top side in a $\sqrt{2}\times\sqrt{2}$ two-dimensional supercell plus a vacuum layer $>$ 10 \AA. The kinetic energy cutoff of the plane-wave basis was chosen to be 400 eV. An 8$\times$8$\times$1 k-point mesh for the Brillouin zone sampling and the Gaussian smearing technique with a width of 0.05 eV were used. The nonmagnetic, checkerboard AFM N\'eel, and collinear AFM states were studied. In all these magnetic orders, the top two slab layers and all atoms in FeSe layer(s) were allowed to relax until the corresponding forces were smaller than 0.01 eV/\AA, while the bottom slab layers were fixed at their bulk positions. The electric field induced by asymmetric slab relaxation was compensated by a dipole correction.\cite{dipole}

We note that as FeSe has charge-neutral Se-Fe$_2$-Se layers, the vdW interaction plays an important role in the interlayer bonding. Our previous studies on bulk FeSe have shown that the lattice parameters (especially along the $c$ direction) and phonon frequencies can be accurately calculated only when the vdW interaction is taken into account.\cite{ye13} In the present study on FeSe epitaxial system, we had checked the equilibrium distance between the top Se atom of the first FeSe layer and the TiO$_2$ termination layer of the substrate. The calculated distance with vdW correction (5.56 \AA) is in excellent agreement with the experimental value (5.5 \AA).\cite{xueCPL12} In comparison, the calculated distance without vdW correction is 5.85 {\AA}. The inclusion of vdW correction in the calculations is very important to get realistic adsorption structural parameters, which is a prerequisite to study the related properties such as the phonon and its derivatives.

After the equilibrium structures were obtained, the frequencies and displacement patterns of phonon modes were calculated using the dynamical matrix method.\cite{liu05} Firstly, we calculated the phonon modes with equilibrium structure in the collinear AFM order (the energetically favorable state). Secondly, even though the ground state is in collinear AFM order, magnetic domains will always form in a real material with magnetic stripe directions randomly along the $a$ or $b$ direction. It turns out that the ARPES measured Fermi surface is an average effect of the electronic states of those magnetic domains. Correspondingly, in calculations, the Fermi surface in the checkerboard AFM N\'eel state can reproduce the ARPES results.\cite{cohen13,zhangp13} Thus we also studied the lattice dynamics in the checkerboard AFM N\'eel order. Our previous study on bulk FeSe has shown that the phonon frequencies calculated using the checkerboard AFM N\'eel order fit the experimental results well. \cite{ye13} We have checked that the different antiferromagnetic orders do not change the phonon frequencies very much ($<$ 2 meV). In total, there were 54 and 78 independent phonon modes calculated for the monolayer and bilayer FeSe epitaxial films, respectively.

Physically the influence of a phonon on the magnetic properties of FeSe epitaxial film happens through the electron-phonon coupling, which has twofold impacts on electronic structure, namely thermally scattering of the electron and quantum zero-point vibration of the phonon. Here we study the impact resulting from the zero-point vibrations of phonon modes. In the calculations, the atomic displacements due to the zero-point vibrations of phonon modes were obtained according to the method of Ref. \onlinecite{west06}. Specifically, the atoms were displaced to a vibrational state with a potential energy of $\hbar\omega_s$/4 instead of $\hbar\omega_s$/2 for each specified phonon mode $s$ according to the Heisenberg uncertainty principle, while its normal-mode coordinates could reach two maxima along two opposite directions.\cite{ye13} When the atoms were displaced, the energies in different AFM orders were calculated on the same distorted structure. The variation of the local magnetic moment on Fe due to zero-point atomic displacement is $|\bigtriangleup M|$ = $|M_+ - M_-|$, with $M_+$ being the local magnetic moment on Fe for one displacement and $M_-$ the other.\cite{ye13}

\begin{table*}[!t]
\caption{The relaxed layer distances $d$ (in \AA) and local magnetic moment $M$ (in $\mu_{\rm B}$) on Fe of monolayer and bilayer FeSe epitaxial films in the collinear AFM order and the checkerboard AFM N\'eel order under different levels of electron doping. The $M_1$ and $M_2$ denote the magnetic moment of Fe in the bottom and top FeSe layers, respectively. The 'N\'eel-Collinear' means the bottom layer of FeSe bilayer is in the checkerboard AFM N\'eel order while the top layer in the collinear AFM order.}
\begin{center}
\begin{tabular*}{17.5cm}{@{\extracolsep{\fill}} cccccccccccccccc}
\hline
\hline
doping(e) & 0.0 & 0.1 & 0.2 & 0.3 & 0.4 & 0.0 & 0.1 & 0.2 & 0.3 & 0.4 & 0.0 & 0.1 & 0.2 & 0.3 & 0.4 \\
\hline
Monolayer & \multicolumn{5}{c}{Collinear} & \multicolumn{5}{c}{N\'eel} \\
\cline{2-6} \cline{7-11}
$d_{\rm TiO-FeSe}$ & 4.246 & 4.246 & 4.249 & 4.248 & 4.249 & 4.189 & 4.194 & 4.199 & 4.200 & 4.200 \\
$M$ & 2.132 & 2.142 & 2.156 & 2.165 & 2.172 & 1.913 & 1.903 & 1.890 & 1.883 & 1.878 \\
\\
Bilayer & \multicolumn{5}{c}{Collinear-Collinear} & \multicolumn{5}{c}{N\'eel-N\'eel} & \multicolumn{5}{c}{N\'eel-Collinear} \\
\cline{2-6} \cline{7-11} \cline{12-16}
$d_{\rm TiO-FeSe}$ & 4.287 & 4.243 & 4.247 & 4.246 & 4.245 & 4.168 & 4.172 & 4.174 & 4.178 & 4.178 & 4.203 & 4.205 & 4.205 & 4.205 & 4.207 \\
$d_{\rm FeSe-FeSe}$ & 5.423 & 5.238 & 5.214 & 5.222 & 5.238 & 5.218 & 5.219 & 5.220 & 5.232 & 5.241 & 5.293 & 5.294 & 5.301 & 5.308 & 5.305 \\
$M_1$ & 2.135 & 2.110 & 2.055 & 2.119 & 2.132 & 1.888 & 1.884 & 1.874 & 1.866 & 1.857 & 1.907 & 1.899 & 1.885 & 1.884 & 1.866 \\
$M_2$ & 2.165 & 2.158 & 2.124 & 2.161 & 2.167 & 1.905 & 1.897 & 1.889 & 1.883 & 1.878 & 2.161 & 2.164 & 2.163 & 2.167 & 2.169 \\
\hline \hline
\end{tabular*}
\end{center}
\end{table*}

Here the electron doping effect can be simulated with either oxygen vacancies in the substrate\cite{zhangsb13,gongxg14} or excess electrons in the system. Considering that the oxygen vacancies may induce some artificial vibrational modes, we thus simulated the electron doping effect by changing the total number of electrons in the system, which reserves the spatial symmetry as the undoped case. As to the amount of doped electrons, the ARPES experiment suggests an electron counting of about 0.1 electrons/Fe by analyzing the area ratio of the Fermi surface around $M$ to the whole Brillouin zone. \cite{zhouxjNC12} Accordingly, we studied the systems with electron doping from 0.1 to 0.4 electrons for the whole supercell. The charge difference density was calculated from $\Delta Q$ = $Q$(FeSe/SrTiO$_3$) - $Q$(SrTiO$_3$) - $Q$(FeSe), where $Q$(SrTiO$_3$) is the charge density of the undoped or electron-doped substrate.

\begin{figure}[!t]
\includegraphics[angle=0,scale=0.32]{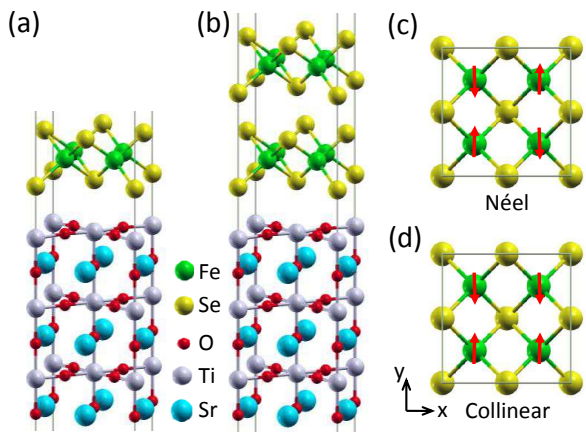}
\caption{(Color online) Atomic structures of (a) monolayer and (b) bilayer FeSe on TiO$_2$-terminated SrTiO$_3$(001) surface. Magnetic patterns of Fe atom spins denoted by red arrows in the (c) checkerboard AFM N\'eel state and (d) collinear AFM state.}
\label{figstruc}
\end{figure}

\section{RESULTS AND ANALYSIS}

The structures and spin patterns of monolayer and bilayer FeSe films on SrTiO$_3$ are shown in Figure \ref{figstruc}. As in the MBE experiment,\cite{xueCPL12} FeSe films are adsorbed on the TiO$_2$-terminated SrTiO$_3$(001) surface. From the total energy calculations, we find that the FeSe epitaxial film energetically favors such adsorption sites at which the bottom Se atoms are on top of the terminated Ti atoms of the substrate [Fig. \ref{figstruc}(a) and \ref{figstruc}(b)]. This is also consistent with our previous calculation.\cite{liu12} For magnetic orders, the nonmagnetic, ferromagnetic, checkerboard AFM N\'eel, and collinear AFM states have been considered. The spin patterns of the latter two cases are shown in Fig. \ref{figstruc}(c) and \ref{figstruc}(d), respectively. The relaxed layer distances $d$ and local magnetic moments $M$ on Fe in the collinear AFM order (the energetically favorable state) and the checkerboard AFM N\'eel order under different levels of electron doping are listed in Table I. For comparison, the structural parameters of the FeSe epitaxial films in the nonmagnetic state are provided in Table II in the appendix.

\begin{figure}[!t]
\includegraphics[angle=0,scale=0.32]{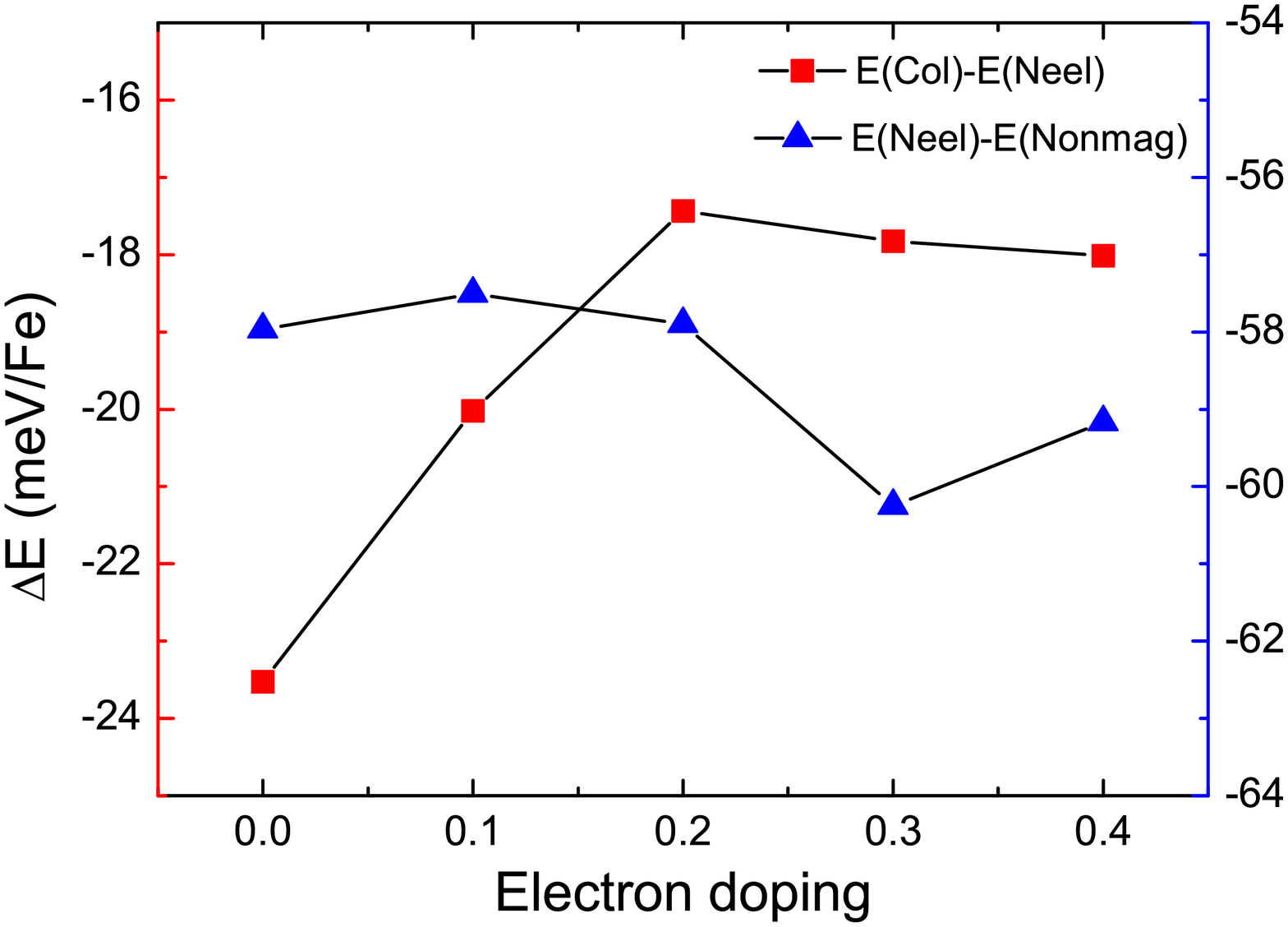}
\caption{(Color online) Energy differences between different magnetic states for monolayer FeSe on TiO$_2$-terminated SrTiO$_3$(001) surface as functions of electron doping.}
\label{fig1Lediff}
\end{figure}

\subsection{Monolayer FeSe film on SrTiO$_3$}

\begin{figure}[!t]
\includegraphics[angle=0,scale=0.42]{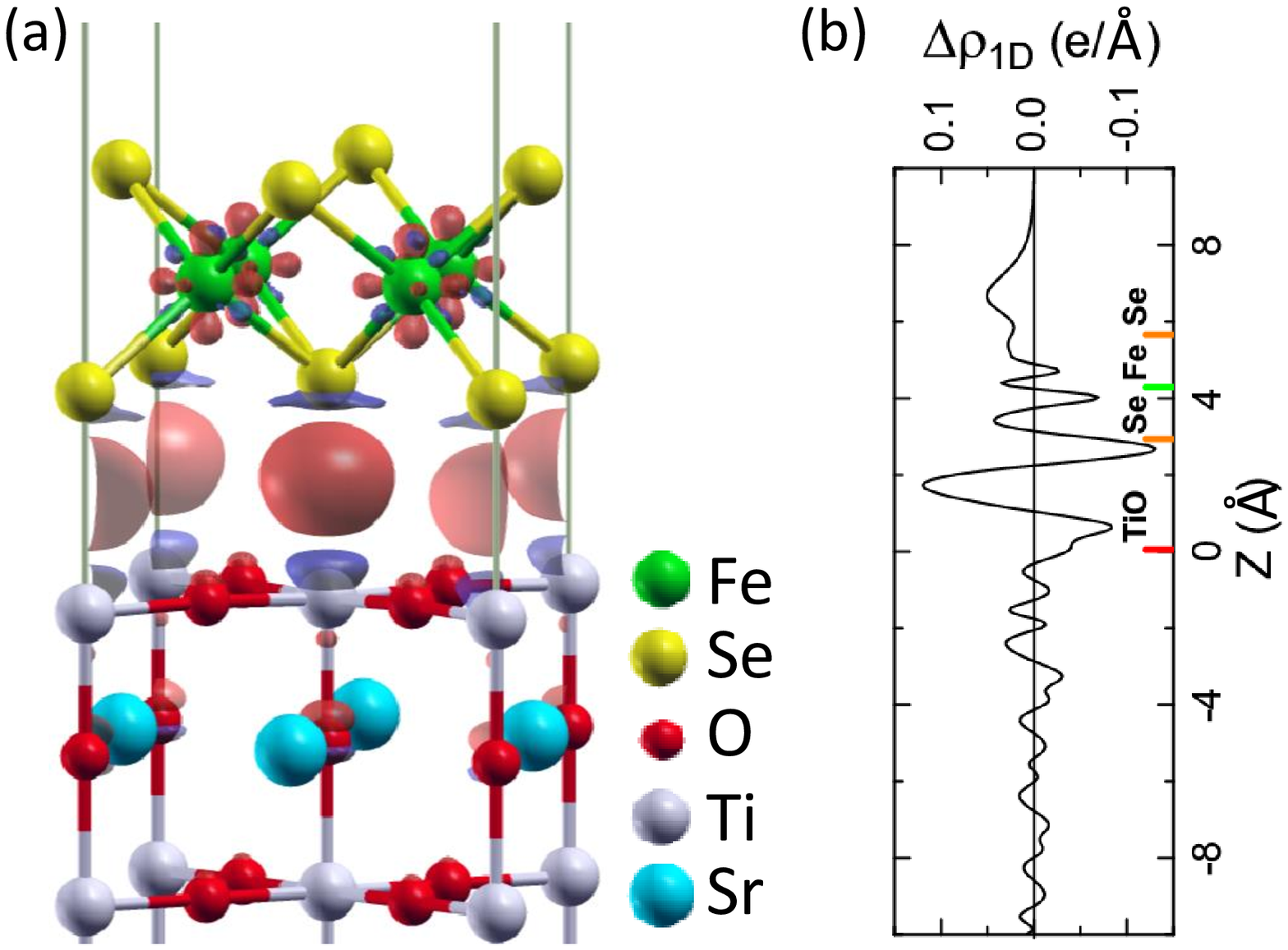}
\caption{(Color online) (a) Three dimensional and (b) one dimensional charge difference density for monolayer FeSe adsorbed on TiO$_2$-terminated SrTiO$_3$(001) surface with 0.2-electron doping. Dark red and blue isosurfaces in panel (a) are respectively electron accumulation and depletion areas in isovalue of 0.01 e/\AA$^3$.}
\label{fig1Lchg}
\end{figure}

\begin{figure}[!b]
\includegraphics[angle=0,scale=0.4]{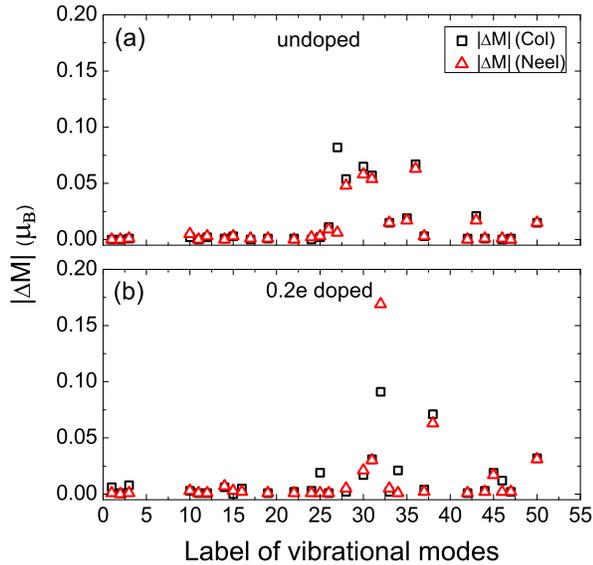}
\caption{(Color online) Variations of local magnetic moment on Fe induced respectively by the zero-point atomic displacements of all the vertical phonon modes for (a) undoped and (b) 0.2-electron doped monolayer FeSe on SrTiO$_3$. The phonon modes calculated at the equilibrium structure of the collinear AFM state.}
\label{fig1LColMvib}
\end{figure}

Figure \ref{fig1Lediff} shows the calculated energy differences between checkerboard AFM N\'eel and nonmagnetic states as well as between collinear AFM and checkerboard AFM N\'eel states as a function of electron doping. As we can see, the checkerboard AFM N\'eel state remains energetically lower than the nonmagnetic state in all the doped cases, and the corresponding energy difference only changes slightly, within 2 meV/Fe. In contrast, the energy difference between the collinear AFM state and the checkerboard AFM N\'eel state decreases considerably when electrons are doped. To be specific, the energy of the collinear AFM state is 23.5 meV/Fe lower than that of the checkerboard AFM N\'eel state in the undoped case, while it decreases to 17.4 meV/Fe in the case of 0.2 electrons doping. However, after more than 0.2 electrons are doped, the energy difference no longer decreases. Thus the electron doping indeed has a notable effect on the magnetic properties of monolayer FeSe epitaxial film, but it is still not enough in and of itself to change the magnetic states energetically.

To clarify how the charge transfer takes place between the doped substrate and the FeSe adlayer, we inspect the calculated charge difference density in the case of 0.2-electron doping, as shown in Fig. \ref{fig1Lchg}. Clearly, the doped electrons converge to the interface between the FeSe adlayer and the SrTiO$_3$ substrate. Meanwhile, the adsorption of the FeSe layer on the substrate induces a certain amount of charge redistributed around the Fe atoms, which is responsible for the energy changes among the different AFM states. Interestingly, a recent study has found strong coupling of the Iron-quadrupole and anion-dipole polarizations in Ba(Fe$_{1-x}$Co$_x$)As$_2$.\cite{mac14} Physically the electron accumulation at the interface provides the bonding interaction between the adlayer and the substrate, which is expected to be very susceptible to the interfacial atomic vibrations.

\begin{figure}[!b]
\includegraphics[angle=0,scale=0.4]{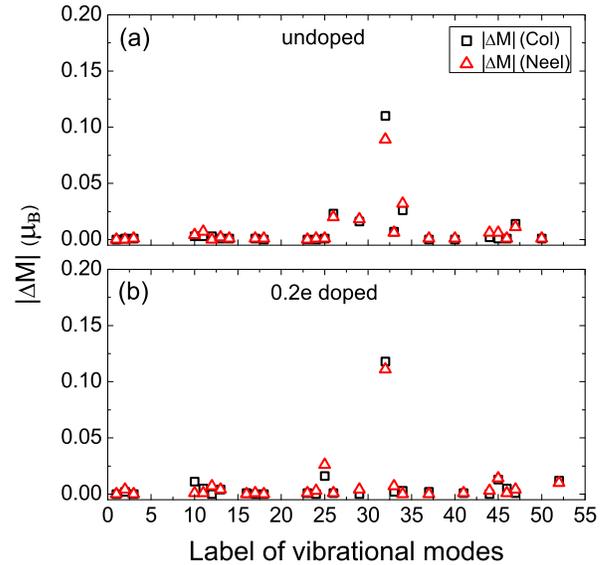}
\caption{(Color online) Variations of local magnetic moment on Fe induced respectively by the zero-point atomic displacements of all the vertical phonon modes for (a) undoped and (b) 0.2-electron doped monolayer FeSe on SrTiO$_3$. The phonon modes calculated at the equilibrium structure of the checkerboard AFM N\'eel state.}
\label{fig1LNeelMvib}
\end{figure}

In our previous study on crystal $\beta$-FeSe under hydrostatic pressure,\cite{ye13} it was found that in comparison with the other phonon modes, the zero-point atomic displacement due to the coherent vertical vibration of Se induces the largest variation of local magnetic moment on Fe.\cite{ye13} This urges us to examine whether or not the same phonon effect exists in FeSe epitaxial films. Figures \ref{fig1LColMvib} and \ref{fig1LNeelMvib} show the calculated variations of the local magnetic moment in monolayer FeSe on SrTiO$_3$ induced, respectively, by the zero-point atomic displacements of all the independent vertical phonon modes. The phonon modes are calculated at the equilibrium structure in the collinear AFM order (Fig. \ref{fig1LColMvib}) and the checkerboard AFM N\'eel order (Fig. \ref{fig1LNeelMvib}), respectively. Only small even negligible variations are found ($<0.2~\mu_B$), regardless of whether the FeSe layer is in a collinear AFM state or in a checkerboard AFM N\'eel state, nor does it matter what the doping level of the electrons is. Here the mode numbered by 32 is exactly the same vibrational mode relating to the Se height from the Fe-Fe plane as found in crystal $\beta$-FeSe in our previous work.\cite{ye13} In comparison, the variation due to the corresponding phonon mode can be as large as 1.2~$\mu_B$ in the case of bulk FeSe under pressure.\cite{ye13}

\begin{figure}[!t]
\includegraphics[angle=0,scale=0.4]{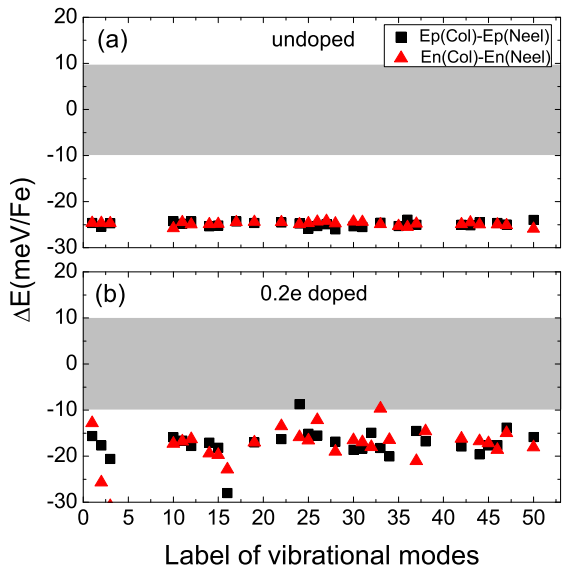}
\caption{(Color online) Energy difference between collinear AFM and checkerboard N\'eel AFM states varying with the positive (p) or negative (n) zero-point atomic displacements of vertical phonon modes for monolayer FeSe on SrTiO$_3$ in (a) undoped and (b) 0.2-electron doped cases. Grey areas highlight the regions with energy difference within $\pm$10.0 meV/Fe. The phonon modes calculated at the equilibrium structure of the collinear AFM state.}
\label{fig1LColEvib}
\end{figure}

\begin{figure}[!t]
\includegraphics[angle=0,scale=0.4]{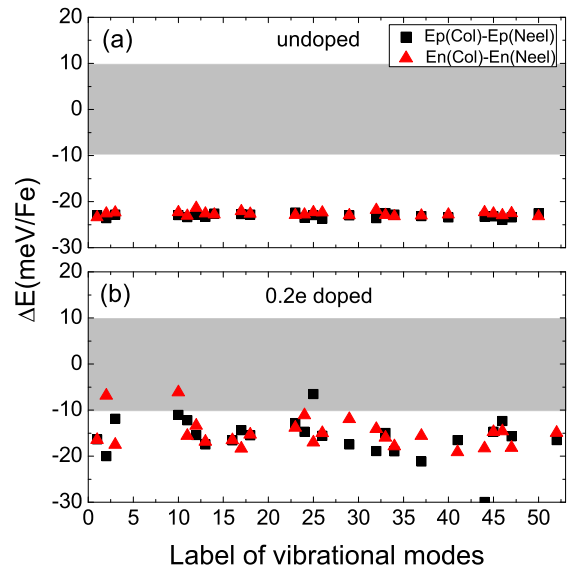}
\caption{(Color online) Energy difference between collinear AFM and checkerboard N\'eel AFM states varying with the positive (p) or negative (n) zero-point atomic displacements of vertical phonon modes for monolayer FeSe on SrTiO$_3$ in (a) undoped and (b) 0.2-electron doped cases. Grey areas highlight the regions with energy difference within $\pm$10.0 meV/Fe. The phonon modes calculated at the equilibrium structure of the checkerboard AFM N\'eel state.}
\label{fig1LNeelEvib}
\end{figure}

Figures \ref{fig1LColEvib} and \ref{fig1LNeelEvib} show how the calculated energy difference between collinear AFM and checkerboard AFM N\'eel states varies with the zero-point atomic displacement of a vertical phonon mode. The phonon modes are calculated at the equilibrium structure in the collinear AFM order (Fig. \ref{fig1LColEvib}) and the checkerboard AFM N\'eel order (Fig. \ref{fig1LNeelEvib}), respectively. For the undoped FeSe monolayer [Figs. \ref{fig1LColEvib}(a) and \ref{fig1LNeelEvib}(a)], we see that the energy difference remains nearly unchanged ($\sim$ -23 meV/Fe) for all the independent vertical phonon modes. In contrast, when 0.2 electrons are doped, dramatic changes occur [Figs. \ref{fig1LColEvib}(b) and \ref{fig1LNeelEvib}(b)]. There are five modes that reduce the energy difference to a value within $\pm$10.0 meV/Fe (represented by the dots in the grey area). As shown in Fig. \ref{figstruc}(c) and \ref{figstruc}(d), the transition between the collinear AFM state and the checkerboard AFM N\'eel state can be viewed as local magnetic moment transfer between Fe atoms or spin flip on Fe. Actually, it is well known that it gives rise to magnetic frustration when the collinear AFM and checkerboard AFM N\'eel states are degenerate energetically.

\begin{figure}[!t]
\includegraphics[angle=0,scale=0.35]{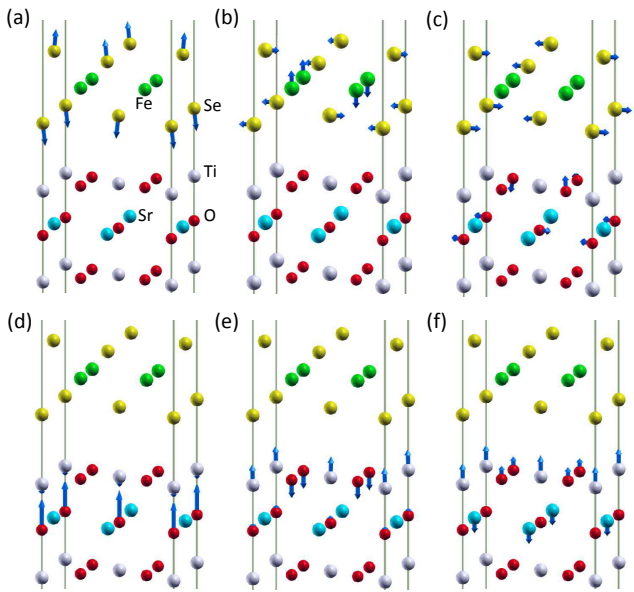}
\caption{(Color online) Atomic displacement patterns for six typical phonon modes [one mode numbered 32 in Figs. \ref{fig1LColMvib}(b) and \ref{fig1LNeelMvib}(b) while five modes dotted in grey region of Figs. \ref{fig1LColEvib}(b) and \ref{fig1LNeelEvib}(b)] of monolayer FeSe on SrTiO$_3$ in 0.2-electron doped case. The arrows represent the direction and amplitude of atomic displacements.}
\label{fig1Lvib}
\end{figure}

The atomic displacement patterns of the above six vibrational modes, among which one induces the largest local moment variation [mode 32 in Figs. \ref{fig1LColMvib}(b) and \ref{fig1LNeelMvib}(b)] and five substantially reduce the energy difference between collinear AFM and checkerboard AFM N\'eel states [dotted in gray region of Figs. \ref{fig1LColEvib}(b) and \ref{fig1LNeelEvib}(b)], are schematically shown in Figure \ref{fig1Lvib}. The first one is the coherent vertical vibrations of Se bridge atoms. The second mode involves vertical vibrations of Fe atoms and lateral vibrations of Se atoms. The following four vibrational modes are contributed by the Ti and O atoms at the interface of FeSe epitaxial film and SrTiO$_3$ substrate. The calculated frequency of the vibrational mode in Fig. \ref{fig1Lvib}(d) is 99 meV, which is very close to an energy difference of 100 meV between two replica electronic bands observed in the ARPES experiment.\cite{shenzx13} The vibrational mode with a frequency of 53 meV shown in Fig. \ref{fig1Lvib}(e) is the ferroelectric phonon.\cite{leedh12,xingdy14} It is the vibrational mode that could enhance the energy scale of Cooper pairing and even change the pairing symmetry, proposed by theorists.\cite{leedh12} We note that the frequency 53 meV of the ferroelectric phonon mode is almost one-half of the frequency 99 meV of the vibrational mode in Fig. \ref{fig1Lvib}(d) and meanwhile is close to the energy distance 50 meV of a weaker replica band from the original band also observed in the ARPES experiment.\cite{shenzx13} In the electron-doped case, these vibrational modes effectively alter the interactions between Fe atoms directly [Fig. \ref{fig1Lvib}(a)-(c)] or indirectly [Fig. \ref{fig1Lvib}(d)-(f)] and affect the magnetic properties of FeSe epitaxial film.

\subsection{Bilayer FeSe film on SrTiO$_3$}

In the STM experiment, it was found that the monolayer FeSe epitaxial film shows a signal of superconducting transition around 77 K while the bilayer FeSe epitaxial film does not show a superconducting gap.\cite{xueCPL12} It is thus strongly desirable to find out the difference between the monolayer and bilayer FeSe epitaxial films. Our previous calculations indicate that the electronic band structures of monolayer and bilayer FeSe on SrTiO$_3$ in the undoped case are very similar.\cite{liu12} In the present study, we devote ourselves to the effects of both electron doping and phonon vibrations on the magnetic properties of FeSe epitaxial films.

\begin{figure}[!t]
\includegraphics[angle=0,scale=0.32]{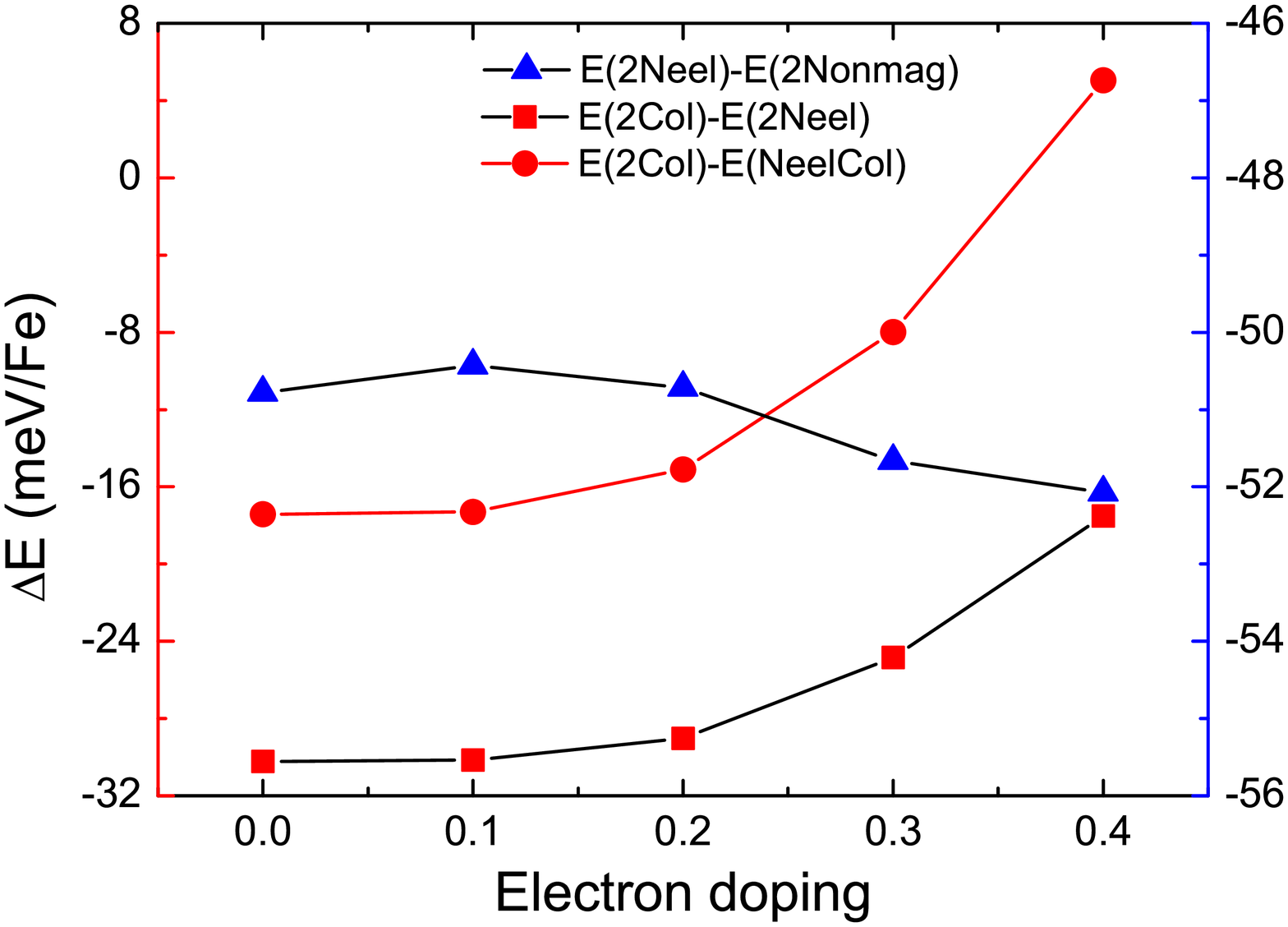}
\caption{(Color online) Energy differences between different magnetic states for bilayer FeSe on TiO$_2$-terminated SrTiO$_3$(001) surface as functions of electron doping.}
\label{fig2Lediff}
\end{figure}

Figure \ref{fig2Lediff} shows the energy differences between different magnetic states for bilayer FeSe on SrTiO$_3$ with and without electron doping. For the bilayer FeSe, we have considered four different magnetic orders, namely i) both layers in nonmagnetic state, ii) both layers in collinear AFM order, iii) both layers in checkerboard AFM N\'eel order, and iv) the bottom layer in checkerboard AFM N\'eel order while the top layer in collinear AFM order. The energy of both layers in checkerboard AFM N\'eel order is always energetically lower than that of the nonmagnetic state. The energy difference between both layers in collinear AFM and both layers in checkerboard AFM N\'eel orders gradually decreases when electrons are doped. However, unlike in the monolayer case (Fig. \ref{fig1Lediff}), the 0.2-electron doping does not change this energy difference substantially in the bilayer case. Actually the energy difference does not decrease to -17 meV/Fe until 0.4 electrons are doped. Overall, comparing Fig. \ref{fig2Lediff} with Fig. \ref{fig1Lediff}, we see that twice as many electrons are required to be doped in bilayer FeSe to induce the same effect as in monolayer FeSe. This agrees with the recent ARPES experiment, which showed the bilayer FeSe film on SrTiO$_3$ is much more difficult to be doped with annealing than the monolayer FeSe.\cite{zhouxj14b} Furthermore, we would like to address another important difference between the bilayer and monolayer FeSe epitaxial films, namely that the bilayer FeSe can adopt different magnetic orders for the bottom and top layers. When the electron doping increases ($>0.35e$), the energetically favorable state transforms from both layers in collinear AFM order into the state of the bottom layer in checkerboard AFM N\'eel order and the top layer in collinear AFM order. This means that enough electron doping alone can induce magnetic frustration in the interface FeSe layer of multilayer FeSe epitaxial films.

\begin{figure}[!t]
\includegraphics[angle=0,scale=0.48]{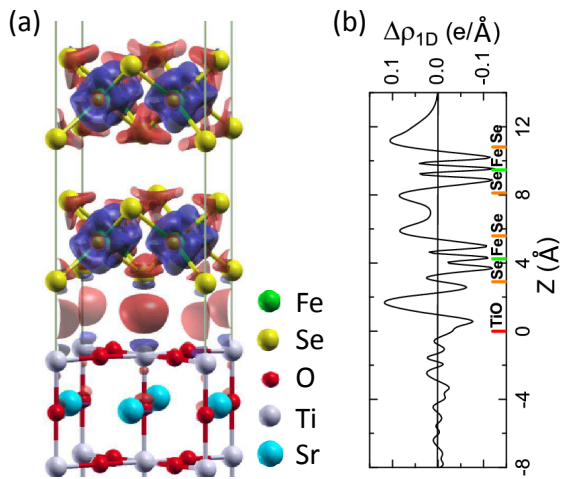}
\caption{(Color online) (a) Three dimensional and (b) one dimensional charge difference density for bilayer FeSe adsorbed on SrTiO$_3$ with 0.2-electron doping. Dark red and blue isosurfaces in panel (a) are respectively electron accumulation and depletion areas in isovalue of 0.01 e/\AA$^3$.}
\label{fig2Lchg}
\end{figure}

In Figure \ref{fig2Lchg}, we plot the calculated charge difference density for bilayer FeSe on SrTiO$_3$. The electron accumulation at the interface is similar to that in the monolayer FeSe epitaxial case (Fig. \ref{fig1Lchg}), meanwhile there is no electron accumulation in the region between the two FeSe layers. Around Fe atoms, there are also charge redistributions. As reported in Tables I, when more than 0.2 electrons are doped in the bilayer FeSe case, the distance between the bottom FeSe layer and the termination TiO$_2$ layer of the substrate changes little. This facilitates the electron doping for the bottom layer in the bilayer FeSe case. On the other hand, the distance between the bottom and top FeSe layers tends to increase when the electron doping exceeds 0.2 e, which makes the top layer hard to be further doped. For the monolayer and bilayer FeSe films, the similar electron accumulation at the interface suggests that the similar physical phenomena would occur at the interfacial FeSe layer.

\section{DISCUSSION}

In iron-based superconductors, it is commonly thought that AFM fluctuations are responsible for the superconductivity.\cite{stewart11,scalapino12} For bulk $\beta$-FeSe under pressure, the zero-point atomic displacement of the vertical vibrational mode of Se atoms induces a large variation on local magnetic moment via strong spin-phonon coupling, as demonstrated in our previous work.\cite{ye13} For the FeSe epitaxial films in the present study, it is found that, unlike bulk FeSe,\cite{ye13} the variation of local magnetic moment via zero-point atomic displacements does not show any meaningful enhancement in both undoped and doped cases (Figs. \ref{fig1LColMvib} and \ref{fig1LNeelMvib}). In contrast, the energy difference between collinear AFM and checkerboard AFM N\'eel states shows a meaningful reduction with electron doping for both the monolayer (Fig. \ref{fig1Lediff}) and bilayer FeSe (Fig. \ref{fig2Lediff}) cases.

For monolayer FeSe on SrTiO$_3$, the dramatic reduction in the energy difference between the collinear AFM and checkerboard AFM N\'eel states is caused by a combined effect of electron doping and phonon vibrations. As shown in Fig. \ref{fig1Lediff}, without considering the phonon effect, the energy difference decreases with electron doping but saturates at about -17 meV/Fe after more than 0.2 electrons are doped. On the other hand, without including electron doping, the absolute value of the minimum energy difference is still larger than 23 meV/Fe when only the phonon effect is taken into account [Figs. \ref{fig1LColEvib}(a) and \ref{fig1LNeelEvib}(a)]. The combined effect of electron doping and phonon vibrations brings about a certain probability to substantially reduce the energy difference [Figs. \ref{fig1LColEvib}(b) and \ref{fig1LNeelEvib}(b)]. In the $J_1$-$J_2$ Heisenberg model,\cite{mafj08} the magnetic coupling $J_1$ between the nearest-neighboring (NN) Fe spins and the $J_2$ between the next-nearest-neighboring (NNN) Fe spins can be derived from the energy differences between different magnetic states. In the undoped case, it gives 29.0 meV/$S^2$ for $J_1$ and 20.8 meV/$S^2$ for $J_2$ with $S$ being the local magnetic moment on Fe. With electron doping and phonon effects, it is very likely that the reduced energy difference between collinear AFM and checkerboard AFM N\'eel states makes $J_2$ comparable with $J_1/2$. Then the magnetic frustration or magnetic instability takes place, which is helpful to superconductivity. This provides a consistent physical picture for previous experimental \cite{xueCPL12,fengdlPRL14,fengdl14,zhouxjNC12,zhouxjNM13,fengdlNM13,xueAPE13,xueCPL14,
maxc14,chucw13,shenzx13,zhouxj14a,zhouxj14b,jiajf14} and theoretical studies on the FeSe monolayer. \cite{liu12,leedh12,cohen13,zhangp13,zhangsb13,kuw14,gongxg14,duanwh14,xingdy14}

The bilayer FeSe epitaxial film behaves distinctly from the monolayer one. At regular doping levels, the Fermi surfaces of monolayer and bilayer FeSe detected in ARPES experiments are quite different.\cite{zhouxjNC12,zhouxjNM13,fengdlNM13} The former shows only electron-type Fermi surfaces around the Brillouin zone corners, while the latter shows both electron-type Fermi surfaces around the corners and hole-type Fermi surfaces around the center of the Brillouin zone. Nevertheless, when it is effectively doped with prolonged annealing time, the bilayer FeSe epitaxial film demonstrates a similar Fermi surface to the monolayer one in ARPES measurement.\cite{zhouxj14b} This means that the top layer of bilayer FeSe is much more difficult to be doped than its bottom layer experimentally. On the other hand, we remind the reader that the Fermi surface of monolayer FeSe observed in ARPES can be well reproduced in calculations when the FeSe layer is set in checkerboard AFM N\'eel order.\cite{cohen13,zhangp13} Meanwhile, our calculations show that to make the checkerboard AFM N\'eel order favorable energetically by doping, the top layer of bilayer FeSe is much harder than its bottom layer (Fig. \ref{fig2Lediff}). This is consistent with the ARPES experiment. \cite{zhouxj14b} In addition, in transport experiments, clear superconducting transitions have been observed both on a five unit-cell FeSe film covered with an amorphous Si protection layer\cite{xueCPL12} and on a monolayer FeSe film protected with FeTe film.\cite{xueCPL14} It has been argued that the superconductivity occurs at the first unit-cell FeSe in transport measurements.\cite{xueCPL12,xueCPL14} In the calculations, we have found that the magnetic frustration in the interfacial FeSe layer of bilayer FeSe film may be induced by electron doping, which supports this viewpoint.


\section{Summary}

We have studied the magnetic properties of monolayer and bilayer FeSe epitaxial films on SrTiO$_3$ with electron doping by using the first-principles calculations with vdW correction. For monolayer FeSe epitaxial film, the combined effect of electron doping and phonon vibrations gives rise to a certain probability to induce magnetic frustration between the collinear AFM and checkerboard AFM N\'eel states, which is helpful to superconductivity. Most of the effective phonon modes, which can substantially reduce the energy difference between AFM states with electron doping, are contributed by interfacial Ti and O atoms. In contrast, for bilayer FeSe epitaxial film, the interfacial FeSe layer has magnetic frustration induced by electron doping much more easily than the top layer, which suggests that for multilayer FeSe epitaxial films, the superconductivity readily takes place at the bottom layer. These calculated results agree with the existing experimental studies.

\begin{acknowledgments}

We wish to thank Qian-Qian Ye and Professor Tao Xiang for helpful discussions. This work is supported by National Natural Science Foundation of China (Grant No. 11004243, No. 11190024, and No. 51271197), National Program for Basic Research of MOST of China (Grant No. 2011CBA00112), the Fundamental Research Funds for the Central Universities, and the Research Funds of Renmin University of China (14XNLQ03). Computational resources have been provided by the Physical Laboratory of High Performance Computing at Renmin University of China. The atomic structures were prepared with the XCRYSDEN program.\cite{kokalj}

\end{acknowledgments}

\appendix*

\section{}

The relaxed structural parameters for monolayer and bilayer FeSe epitaxial films in the nonmagnetic state are listed in Table II.

\begin{table}[h]
\caption{The relaxed layer distances $d$ (in \AA) of monolayer and bilayer FeSe epitaxial films in the nonmagnetic state under different levels of electron doping.}
\begin{center}
\begin{tabular*}{8.0cm}{@{\extracolsep{\fill}} cccccc}
\hline
\hline
doping(e) & 0.0 & 0.1 & 0.2 & 0.3 & 0.4 \\
\hline
Monolayer \\
$d_{\rm TiO-FeSe}$ & 4.138 & 4.134 & 4.137 & 4.137 & 4.135 \\
\\
Bilayer \\
$d_{\rm TiO-FeSe}$ & 4.154 & 4.138 & 4.143 & 4.140 & 4.134 \\
$d_{\rm FeSe-FeSe}$ & 5.098 & 5.127 & 5.127 & 5.132 & 5.146 \\
\hline \hline
\end{tabular*}
\end{center}
\end{table}

\end{document}